\documentstyle[prl,aps,twocolumn,psfig]{revtex}
\draft
\begin{document}

\title{A Diffusion-Based Approach to Geminate Recombination of Heme Proteins 
with Small Ligands}

\author{V. S. Starovoitov$^1$ and B. M. Dzhagarov$^2$}

\address{$^1$ B.I.Stepanov Institute of Physics, NASB, 220072, Scarina ave. 70, Minsk,
Belarus \\
$^2$ Institute of Molecular and Atomic Physics NASB, 220072, Scorina ave. 70, Minsk,
Belarus }

\date{\today}

\maketitle
\begin{abstract}
A model of postphotodissociative monomolecular (geminate) recombination of
heme proteins with small ligands ($NO$, $O_2$ or $CO$) is represented. The
non-exponential decay with time for the probability to find a heme in 
unbound state is interpreted in terms of diffusion-like migration of ligand
between protein cavities. The temporal behavior for the probability is
obtained from numerical simulation and specified by two parameters: the time 
$\tau $$_{reb}$ of heme-ligand rebinding for the ligand localized inside the
heme pocket and the time $\tau $$_{esc}$ of ligand escape from the pocket.
The model is applied in the analysis of available experimental data for
geminate reoxygenation of human hemoglobin $HbA$. Our simulation is in good
agreement with the measurements. The analysis shows that the variation in $%
pH $ of the solution ($6.0<pH<9.4$) results in considerable changes for $%
\tau $$_{reb}$ from 0.36 ns (at $pH=8.5$) up to 0.5 ns ($%
pH=6.0$) but effects slightly on the time $\tau $$_{esc}$ ($\tau
_{esc}\approx 0.88$ ns).
\end{abstract}

\narrowtext

\section{Introduction}

The binding reactions between myoglobin ($Mb$) or hemoglobin ($Hb$) and
small ligands ($NO$, $O_2$ or $CO$) are objects of extensive investigations
during many decades because of the great functional importance of the heme
proteins for living systems \cite{ref1,ref2,ref3}. In the investigations a
special attention is paid to the heme-ligand recombination process going
after fast photodissociative bond breaking between a ligand molecule and an
ion $Fe++$ located in the center of a heme ($Fe$-protoporphyrin IX complex).
The kinetic study of the postphotodissociative recombination allows to
obtain detailed information on the protein-ligand interaction mechanism, the
protein structure, the allosteric effect and the medium influence on the
recombination efficiency (see, for example, references \cite
{ref4,ref5,ref6,ref7,ref8,ref9}).

The heme is well-wrapped in protein helixes, which prevent the iron from the
solvent and hinder the ligand migration through protein matrix. On a
sufficiently long time scale (at $t\leq 100$ ns in the case of $Hb$) after
dissociation, when the ligand is not managed to leave the protein and to
move significantly away from the parent heme, the recombination is a
monomolecular reaction designated usually as a geminate recombination (GR) 
\cite{ref5}. Schematically, the GR can be written as \cite{ref10}: 
\begin{equation}
\label{eq1}A\stackrel{k_{reb}}{\longleftarrow }B\stackrel{k_{esc}}{%
\longleftrightarrow }\left\{ C_1,\ldots ,C_n\right\} 
\end{equation}
where $A$ is the bound heme-ligand state. The substates $B$ and $\left\{
C_1,\ldots ,C_n\right\} $ form the unbound state. Each of these substates
corresponds to ligand localization in an individual cavity of protein. The
substate $B$ answers to the residence of ligand inside the heme pocket (the
cavity nearest to the iron on the distal side of heme). The rate constants $%
k_{reb}$ and $k_{esc}$ specify two competing processes: the irreversible
heme-ligand rebinding for the ligand localized inside the heme pocket (that
is, the transition from the substate $B$ to the state $A$) and the migration
of unbound ligand between the heme pocket and other protein cavities (the
transitions between $B$ and the substates $\left\{ C_1,\ldots ,C_n\right\} $%
). Immediately after photodissociation the unbound ligand is 
in the substate $B$. Therefore the
quantity $k_{esc}$ can be associated with ligand escape from the pocket. In
general, the GR is essentially determined by the specificities of
heme-ligand interaction (including the spin restriction effect, 
the position and the
orientation of ligand with respect to the heme plane) \cite
{ref5,ref11,ref12,ref13} and the effect of residues surrounding the heme 
\cite{ref14,ref15,ref16,ref17,ref18,ref19}. Important factors for the
heme-ligand rebinding are also the state of tertiary \cite
{ref2,ref20,ref21,ref22} and quaternary \cite{ref2,ref23,ref24,ref25}
structures of protein, the conformation transitions in protein \cite
{ref2,ref10,ref26,ref27,ref28} and the solvent impact \cite
{ref23,ref29,ref30,ref31}. As a consequence, the kinetic curve
(that is, the probability $P\left( t\right) $ to find the heme in 
unbound state) of GR is a non-exponentially decaying function 
of time \cite{ref32,ref33,ref34,ref35}. After realization of the geminate stage 
a portion 
$P_s$ of the hemes remains in unbound state: $P_s\leq 0.01$ for $NO$, $%
P_s\sim 0.1\div 0.2$ for $O_2$ and $P_s\sim 0.5\div 1.0$ for $CO$. The
quantity $P_s$ characterizes the efficiency of ligand escape from the
protein to the solvent.

Molecular dynamics simulations \cite{ref11,ref36,ref37,ref38,ref39,ref40}
show that the movement of unbound $NO$, $O_2$ or $CO$ ligands in heme
protein can be associated both with ligand trapping for a significant time
in individual cavities and with rare jump-like transitions between adjacent
cavities. It implies a fast establishment of equilibrium for the probability
distribution of ligand within individual cavities. The establishment occurs
on a time scale comparable to the mean time interval $\tau _w$ between the
collisions of ligand with cavity walls. At room temperatures the time $\tau
_w$ lies in the subpicosecond range ($\tau _w\sim 0.1$ ps for $NO$ in the
heme pocket of $Mb$ \cite{ref41}). The ligand redistribution between protein
cavities is observed on a longer time scale ranging from several tens of
picoseconds ($\sim 40$~ps for $NO$ in $Mb$ \cite{ref41}) up to several tens
of nanoseconds ($\sim 50$~ns for $CO$ in $Hb$ \cite{ref42}). Unfortunately,
in practice the detailed molecular dynamics simulation can not be
implemented to the GR due to enormous computational efforts.

In the study we apply an alternative approach based on the diffusion
approximation to ligand migration in protein. Such an approximation is
valid for times $t\gg \tau _w$ when the deterministic nature of ligand
motion can be ignored. Here the interval $\tau _w$ can be recognized as a
correlation time. The diffusion-like character of ligand migration in the
heme proteins can be a reason of the non-exponential temporal dependence for
the probability $P(t)$ \cite{ref43,ref44,ref45}. For instance, a
two-dimensional diffusion is demonstrated for $CO$ in $Mb$ \cite{ref45} to
explain the power-law kinetics to be observed in the experiment. Generally,
reaction (\ref{eq1}) can be represented in a three-dimensional diffusion
approximation by equation 
\begin{equation}
\label{eq2}\frac{\partial n}{\partial t}={\bf \nabla }\left( D{\bf\nabla }%
n\right) -R_{reb}n
\end{equation}
with the diffusion coefficient $D=D(x,y,z)$. The quantity $n=n(x,y,z,t)$ is
the probability density of unbound ligand in the protein. The stepwise
function $R_{reb}=R_{reb}(x,y,z)$ specifies the heme-ligand rebinding and
equals to $k_{reb}$ inside the heme pocket or to zero otherwise.

In order to solve diffusion equation (\ref{eq2}) and to follow the evolution
of GR we use a simple model proposed recently in \cite{ref46}. The model
reproduces dynamics of random walk of particle in porous media 
(such, for instance, as glass-like matrices \cite{ref47,ref48,ref49})
and takes 
into account an initial retention of ligand inside the heme pocket (that is,
in the substate $B$).  In the absence of heme-ligand
rebinding the substate $B$ is realized at times $t<\tau _{esc}$ ($\tau
_{esc}=1/k_{esc}$ is the time of ligand escape from the heme pocket to
others cavities). Only on a longer time scale ($t>\tau _{esc}$) the ligand
succeeds to leave the pocket and to migrate over protein cavities. Due to
the diffusion nature of the migration the time $\tau _{esc}$ can be
specified in terms of the diffusion coefficient $D$.

The approach is implemented with the help of a numerical simulation where
the unbound ligand is represented by a structureless particle. For
simplicity, in the simulation we make some assumptions. The ligand migration
is assumed to be restricted to the distal side of heme. The ligand motion
(realized on a short time scale $t\leq \tau _w$) inside the heme pocket is
represented by a unforced displacement of the particle within a restricted
hemispheric region of space. At $\tau _w\ll t\ll \tau _{esc}$ the ligand
trajectories are effectively mixed in the configurational space, resulting
in a homogeneous distribution for the ligand inside the cavities. Hence, the
probability of irreversible heme-ligand rebinding is accepted to be uniform
for the whole heme pocket. We take into account also that on the time
scale $t\gg \tau _w$ the fast intracavity displacements of ligand for the
substates $\left\{ C_1,\ldots ,C_n\right\} $ do not influence essentially on
the GR kinetics and can be ignored in the simulation. Therefore the ligand
displacement exterior to the heme pocket is simulated as a random walk (that
is, as a Brownian-like motion) of the particle outside the hemispheric
region. This walk is a spatially homogeneous diffusion with the diffusion 
coefficient $D$. We neglect also the structural transformations
(such as a shift of the iron with respect to the porphyrin ring plane) at
the conformational transition of protein between the unliganded and liganded
states. According to the model, the temporal behavior for the probability $%
P(t)$ can be specified in terms of two parameters: the time $\tau _{esc}$
and the time $\tau _{reb}=1/k_{reb}$ of heme-ligand rebinding. The
description of the model is represented in Section 2.

In order to demonstrate the usefulness of such an approach to the GR of heme
proteins we apply the model to the analysis of available experimental data.
We analyze the measured recombination kinetics and the efficiency for a
postphotodissociative GR of human hemoglobin $HbA$ \cite{ref50,ref51}. These
measurements were carried out at various $pH$ values of the solution. Here
we determine the times $\tau _{reb}$ and $\tau _{esc}$ as functions of $pH$ 
and estimate the
influence of solution properties on the heme-oxygen rebinding, the migration
of oxygen molecule in hemoglobin and the efficiency of oxygen escape from
the protein. The association of the times $\tau _{reb}$ and $\tau _{esc}$
with the time of a bimolecular recombination process for hemoglobin is
analyzed. The results of simulation and their analysis are represented in
Section 3.

\section{Diffusion-based model of geminate recombination}

The movement of unbound ligand is considered in a Cartesian coordinate
system $xyz$ attached rigidly to the heme group of atoms. The system origin
is superposed on an iron atom located in the middle of heme porphyrin
ring. The $x$ and $y$ axes are aligned with the heme plane. The positive
direction for the $z$ axis corresponds to the distal side of heme. The
ligand migration in protein is simulated as a probability redistribution for
the ensemble of structureless particles over a three-dimensional
hemispheric space with $z>0$. As in \cite{ref52}, in our simulation the
heme pocket is represented by a hemispheric region 
(designated here as a cage) of radius $\rho $. 
At an initial time instant the particle is uniformly distributed 
inside the cage.

The individual particle to be exposed to a sequence of $\delta $-shaped
uncorrelated kicks executes a random walk in the space. As for the Brownian
particle, each kick results in an abrupt change in the particle velocity.
Between the kicks the particle is in unforced motion. On a time interval $%
\Delta t_k=t_{k+1}-t_k$ ($t_k$ is the time instant of action for $k$-th
kick) between adjacent kicks the particle is specified by the velocity ${\bf %
v}_k$ and the length $L_k$ of free path (note that $\Delta t_k=L_k/\left| 
{\bf v}_k\right| $ ). Then the radius vector ${\bf r}(t_{k+1})$ of particle
for the time point of $k+1$-th kick can be obtained from iteration procedure 
\begin{equation}
\label{eq3}{\bf r}\left( t_{k+1}\right) ={\bf r}\left( t_k\right) +\frac{L_k%
{\bf v}_k}{\left| {\bf v}_k\right| }
\end{equation}
where the radius vector ${\bf r}(t_k)$is given for the time instant 
of $k$-th kick. The projections 
$v_{j,k}$ ($j=x,y,z$) of the velocity ${\bf v}_k$ onto the coordinate axes
and the length $L_k$ are accepted to be independent random quantities, new
values of which are generated at each kick. The quantities $v_{j,k}$ is
obtained from the Maxwell distribution
\begin{equation}
\label{eq4}P_M\left( v_{j,k}\right) =\sqrt{\frac m{2\pi kT}}\exp \left( -
\frac{mv_{j,k}^2}{2kT}\right) 
\end{equation}
Here $m$ is the particle mass and $T$ is a protein temperature. At an
attainment of the $z=0$ plane bounding the space, a new particle velocity
with $v_z>0$ is regenerated in accordance with distribution (\ref{eq4}).

The choice of free path length is dictated by the particle location in the
space. Within the hemispheric cage the particle displacement is unforced and
the particle undergoes no kicks. The length $L_k$ is determined then from
the ballistic trajectory of particle between the cage boundaries. In this
case the length is comparable to the cage size $\rho $. We accept here that
the mean time $\tau _h=\langle \Delta t_{h,k}\rangle $, during which the
particle crosses the cage, can be associated with the time interval $\tau _w$
between the collisions of ligand with heme pocket walls: $\tau _h\sim \tau _w
$.

Exterior to the cage, the particle is exposed to uncorrelated kicks. The
absence of correlation between the kicks implies that the quantity $L_k$ is
distributed according to the exponential law:
\begin{equation}
\label{eqexp}P\left( L_k\right) =\frac 1\lambda \exp \left( -\frac{L_k}%
\lambda \right) 
\end{equation}
where $\lambda =\langle L_k\rangle $ is the mean length of free path for the
particle displacement outside of the cage. The mean time $\tau _c$ between
adjacent kicks and the length $\lambda $ are related to the diffusion
coefficient $D=\langle L_k^2\rangle /6\tau _c$ by equations:
\begin{equation}
\label{eq5}\tau _c=\frac{6mD}{\pi kT}
\end{equation}
\begin{equation}
\label{eq6}\lambda =3D\sqrt{\frac{2m}{\pi kT}}
\end{equation}

Thus, the spatial displacement of the particle is obtained from iterative
equation of motion (\ref{eq3}) and depends on the random sampling of
variables $v_{x,k}$, $v_{y,k}$, $v_{z,k}$ and $L_k$, the statistical
distributions for which are specified by the parameters $m/T$, $D$ and $\rho 
$. As mentioned above, under the conditions typical for the heme proteins
(that is, the temperature, the ligand mass and the distinctive sizes of heme
pocket) the times $\tau _c$ and $\tau _h$ to be accepted here as correlation
times are negligibly short as compared to the characteristic times of GR.
The length $\lambda $ is essentially small as against the size $\rho $ of
hemispheric cage. Hence, the temporal behavior for the probability
redistribution of ligand in heme protein can be described in
terms of the diffusion-based approach.

\begin{figure}
\psfig{file=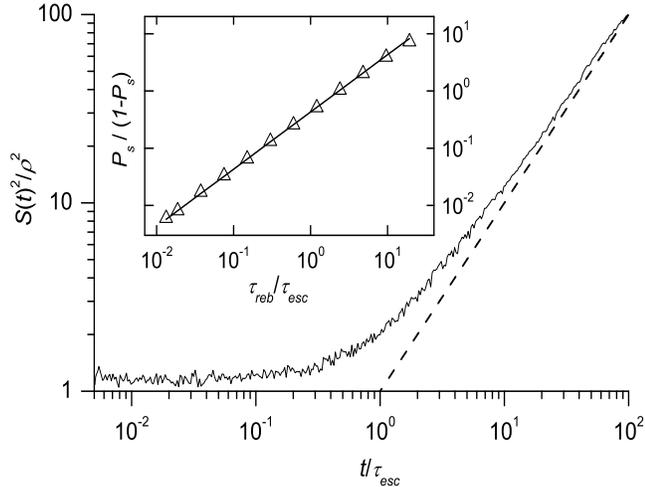,width=8.5cm,height=6.5cm}
\caption{A simulated temporal dependence for the mean-square relative displacement 
$S(t)^2/\rho ^2$ of particle from the initial position at $m=32$ amu, $T=300$
K, $\rho =4$ \AA, $D=3.2\cdot 10^{-11}$ m$^2$/s (solid line). The dashed
line shows dependence $S(t)^2/\rho ^2=t/\tau _{esc}$ corresponding to the
diffusion law. The inset shows the quantity $Ps/(1-Ps)$ 
as a function of the ratio $\tau _{reb}/\tau _{esc}$. 
($\Delta $ - our simulation at $m=32$ amu, $T=300$ K, $\rho =4$ \AA ). 
The solid line gives approximation (\ref{eq8}) at $C_s=0.43$.}
\label{fig1}
\end{figure}

Our model reproduces dynamics of ligand migration over protein cavities. 
Initially, the ligand is retained inside the heme pocket and 
the root-mean-square displacement 
$S(t)=\sqrt{\langle\left| {\bf r}(t)-{\bf r}(0)\right| ^2\rangle }$ 
of ligand from the initial position does not exceed the characteristic size 
of the pocket. In a sense such a retention is analogous to the so called 
cage-effect to be observed for single atoms or small molecules 
in porous glass-like matrices \cite{ref47,ref48,ref49}. 
The time scale, on which the retention is realized, is limited 
by a time point $\tau _{esc}$. This time is a lifetime for the ligand 
inside the heme pocket in the absence of rebinding and 
specifies thereby a ligand escape from the pocket. 
Only on a longer time scale (when the ligand succeeds 
to leave the pocket and to migrate over the protein) 
the ligand displacement $S(t)$ starts to increase significantly. 
According to the model, we associate the time $\tau _{esc}$ 
with the time of particle localization in the hemispheric cage. 
In the simulation the particle displacement $S(t)$ does not exceed 
the cage radius $\rho $ on the short time scale $t<\tau _{esc}$. 
At longer times the quantity $S(t)$ increases with time. 
Due to the diffusion nature of the particle displacement 
the increase in $S(t)^2$ is a linear function of time and 
$S(t)^2\approx 6Dt$ at $t\gg \tau _{esc}$. The relation between 
the time $\tau _{esc}$ and the diffusion coefficient $D$ 
can be then determined from the requirement 
$\rho ^2\sim S(\tau_{esc})^2=6D\tau _{esc}$:
\begin{equation}
\label{eq7}\tau _{esc}=\frac{\rho ^2}{6D}
\end{equation}
Fig. \ref{fig1} demonstrates a typical temporal dependence for
the relative particle displacement 
$S(t)^2/\rho ^2$ simulated within the framework of our model.
The displacement $S(t)$ is shown in the figure to be constant ($S(t)\sim
\rho $) at $\tau _h\ll t\ll \tau _{esc}$. On a longer time
scale ($t\gg \tau _{esc}$) the quantity $S(t)$ approaches
asymptotically the diffusion law: $S(t)^2\approx 6Dt=t\rho ^2/\tau _{esc}$.
Notice that for the time scale $\tau _h\ll t\ll \tau _{esc}$ the temporal
behavior of relative displacements $S(t)/\rho $ is specified by the only
parameter $\tau _{esc}$. In the following, we will adjust the parameter $%
\tau _{esc}$ in the simulation. For definiteness, this adjustment will be
carried out by means of variation in the diffusion coefficient $D$. The
particle mass $m$, the temperature $T$ and the cage radius $\rho $
will take fixed values typical for the ligand and the protein.

The heme-ligand rebinding is accepted to be an irreversible process
occurring when the ligand is localized inside the heme pocket. Therefore
this process is simulated as a random 'death' for the particle within the
hemispheric cage. The particle with $\left| {\bf r}(t_k)\right| <\rho $ is
'obliterated' if $\xi _k\leq \Delta t_{h,k}/\tau _{reb}$. Here $\xi _k$ is a
random quantity to be generated for each period $\Delta t_{h,k}$ when the
particle crosses the cage. The quantity $\xi _k$ is distributed uniformly in
the interval $[0,1]$. The 'obliterated' particle is excluded from the
following consideration.

The probability $P(t)$ to find the heme in unbound state is found as the
ensemble-averaged relative number of the 'non-obliterated' particles at a
time instant $t$. In contrast to the relative displacement $S(t)/\rho $, the
temporal behavior for the probability $P(t)$ depends not only on the
diffusion properties, but on the rate of heme-ligand rebinding as well.
Hence, the behavior of $P(t)$ can be specified in terms of the times $\tau
_{esc}$ and $\tau _{reb}$. In general, the probability $P(t)$ is a
monotonously decreasing function of time, which approaches asymptotically a
steady value $P_s$ at $t\rightarrow \infty $. This value gives a portion of
the hemes remaining in unbound state after realization of GR. As in
diffusion equation (\ref{eq2}), in our model the quantity $P_s$ is a
function dependent merely on the ratio between $\tau _{esc}$ and $\tau _{reb}
$. The analysis of simulated data shows that for a wide range of values $%
\tau _{esc}$ and $\tau _{reb}$ satisfying the requirement $\tau _{reb}/\tau
_{esc}<20$ (that is, under conditions typical for the GR) the best
approximation of the dependence can be represented by relation
\begin{equation}
\label{eq8}\frac{P_s}{1-P_s}\approx C_s\frac{\tau _{reb}}{\tau _{esc}}
\end{equation}
where the coefficient $C_s$ is obtained from mean square fitting. At $m=32$
amu, $T=300K$ and $\rho =4\AA $ the fitting gives a value $C_s=0.43$. The
inset of Fig. \ref{fig1} demonstrates a good agreement between 
approximation (\ref{eq8}) and the simulated data.

\section{Geminate recombination of human hemoglobin with oxygen}

We use the described model in order to analyze available experimental data
for a postphotodissociative reoxygenation of human hemoglobin $HbA$. The
data include the measured recombination kinetics and the efficiency of
oxygen escape from the protein to the solvent for the monomolecular
(geminate) and bimolecular stages of recombination reaction
\begin{equation}
\label{eq9}Hb\left( O_2\right) _3+O_2\longrightarrow Hb\left( O_2\right) _4
\end{equation}
going at room temperatures after fast laser-initiated breaking of a $Fe-O_2$
bond \cite{ref51,ref51}. The kinetic measurements are carried out with a
time resolution of 10 ps for a time scale $t<1.5$ ns at different fixed $pH$
values for the solution. The values of $pH$ fall within an interval
between 6.0 and 9.4: $pH=\left\{ 6.0,6.8,7.0,7.2,7.7,8.0,8.5,9.4\right\} $.

According to the model, 
in the analysis of reoxygenation reaction (\ref{eq9}) the temporal decay for
the probability $P(t)$ is interpreted as a result of two competing
processes: the heme-oxygen rebinding for the oxygen molecule localized
inside the heme pocket and the diffusion-like migration of the oxygen
between hemoglobin cavities. Here we determine the times $\tau _{reb}$ and $%
\tau _{esc}$, which specify the processes. We determine the times as
functions of $pH$ and analyze the effect of solution properties on the
processes to be considered. Due to the tetramer arrangement of hemoglobin
(the $Hb$ molecule consists of heme containing $\alpha $- and $\beta $%
-chains) the observed kinetic curve represents a reoxygenation kinetics
summarized over the chains. Here we make no distinction for reaction (\ref
{eq9}) between the $\alpha $- and $\beta $-chains and determine thereby
chain-averaged times.

\subsection{Reoxygenation kinetics for hemoglobin}

The analysis of reoxygenation kinetics for the hemoglobin is based on the
estimation of the times $\tau _{reb}$ and $\tau _{esc}$. 
The times are found
with the help of a numerical simulation, the iterative procedure for which
is described above (see Section 2). In the simulation the masse of
walking particle is accepted to equal the mass of oxygen molecule. The
temperature $T$ is $300K$. The size $\rho $ of hemispheric cage 
is $4$ \AA
that
corresponds to the time $\tau _h<1$ ps. The times $\tau _{reb}$ and $\tau
_{esc}$ are chosen from an interval of values from 0.1 up to 5 ns. The
correlation time $\tau _c$ and the mean length $\lambda $ of free path are
determined by relations (\ref{eq5}) and (\ref{eq6}). They are negligibly
small in comparison with $\tau _{reb}$, $\tau _{esc}$ or $\rho $. The
simulated dependences for the probability $P(t)$ are obtained from ensemble
averaging for more than $10^6$ particles.

\begin{figure}
\psfig{file=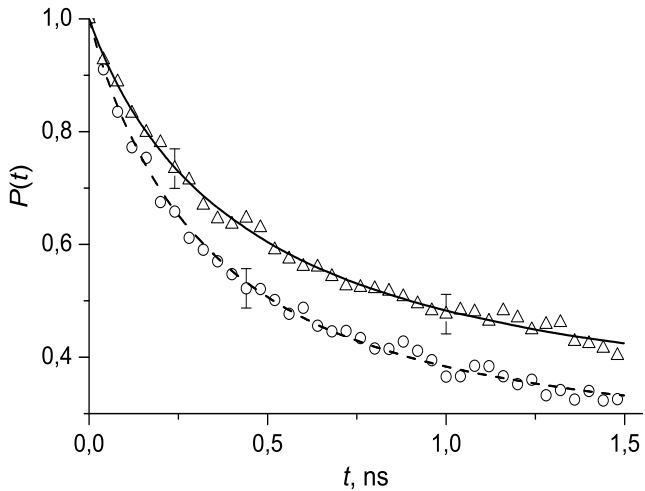,width=8.5cm,height=6.5cm}
\vspace{1 mm}
\caption{A temporal dependence for the probability $P(t)$ to find the
heme in unbound state after fast breaking of a $Fe-O_2$ bond
in human hemoglobin $HbA$ for the geminate stage of reaction (\ref{eq9})
at $pH=6.0$ (solid line - our simulation with $\tau_{reb}=0.495$ ns and 
$\tau _{esc}=0.84$ ns, $\Delta $ - experiment \protect\cite{ref50}) 
and 8.5 (dashed line - our simulation with $\tau _{reb}=0.366$ ns and 
$\tau _{esc}=0.92$ ns, $\circ $ - experiment \protect\cite{ref50}) .}
\label{fig2}
\end{figure}

In the simulation the parameters $\tau _{reb}$ and $\tau _{esc}$ are so
adjusted that the ensemble-averaged temporal dependence of simulated
probability $P(t)$ is the best agreement with a measured kinetic curve.
The agreement is specified by the relative root-mean-square deviation $R$
between the simulated and experimental curves. The simulated dependence for $%
P(t)$ is shown in Fig. \ref{fig2} to reproduce well kinetic
measurements on the considered time scale. The minimal deviation $R$
achieved in our calculations for each of the fixed $pH$ values does not
exceed the measurement error ($R\leq 0.01$). Such an agreement testifies
that the non-exponential dependence for $P(t)$ with time can be explained by
a diffusion-like migration of ligand over protein matrix. Hence, the
parameters $\tau _{reb}$ and $\tau _{esc}$ can be used for the analysis of
the processes, which are responsible for the GR.

\begin{figure}
\psfig{file=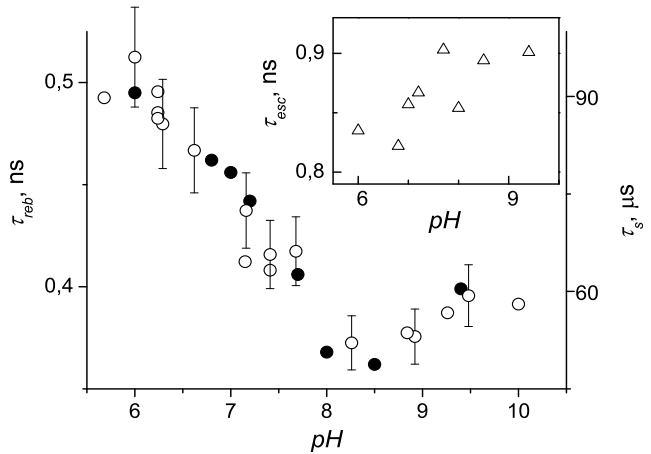,width=8.5cm,height=6cm}
\vspace{1 mm}
\caption{
The times of heme-oxygen rebinding as functions of $pH$ for the
geminate ($\tau _{reb}$, full circles - our simulation) and bimolecular 
($\tau _s$, $\bigcirc$ - experiment \protect\cite{ref50,ref51}) stages 
of recombination reaction (\ref{eq9}). 
The inset shows the $pH$-dependence ($\Delta $) obtained in our simulation 
for the time $\tau _{esc}$ of oxygen
escape from the heme pocket of $HbA$.}
\label{fig3}
\end{figure}

The influence of solution properties on the migration and the rebinding of
oxygen molecule in hemoglobin is assessed from a $pH$ dependence 
for the obtained
times $\tau _{reb}$ and $\tau _{esc}$. Our simulation demonstrates a
significant variation in the rate of heme-oxygen rebinding with $pH$ (see
Fig. \ref{fig3}). The increase of quantity $pH$ from 6.0 to 8.5 
results in a shortening for the time $\tau _{reb}$ by a factor
of 1.4 (from 0.5 down to 0.36 ns). With the following rise of $pH$ to 9.4 
the parameter $\tau _{reb}$ appears to increase up to 0.4 ns. The minimum
magnitude of $\tau _{reb}$ is observed at $pH=8.5$. Despite the considerable 
$pH$ effect for the heme-oxygen rebinding, the variation in $pH$ influences
slightly on the oxygen escape from the heme pocket. The time $\tau _{esc}$
is shown in Fig. \ref{fig3} to be within a range of values from 0.82 to
0.92 ns and to be weakly dependent on $pH$. The $pH$-averaged magnitude of $%
\tau _{esc}$ is approximately 0.88 ns. Notice that this magnitude is larger
than $\tau _{reb}$ by a factor of $2\div 3$.

The obtained values for the times $\tau _{reb}$ and $\tau _{esc}$ are in
good agreement with the experimental study of the alkaline Bohr effect
(the variation of the recombination rate $1/\tau _{s}$ 
with $pH$ for the bimolecular stage of
GR) \cite{ref50,ref51}. The behavior of $pH$-dependence for the time $\tau
_{reb}$ is demonstrated in Fig. \ref{fig3} to be similar to one for the
time $\tau _s$ of bimolecular rebinding. Such an agreement testifies that
for the monomolecular GR the variation of the rebinding rate with $pH$ can
be associated with the same structural transformation as for the bimolecular
stage of reaction (\ref{eq9}). Histidine imidazoles of $C$-terminal sites
and $\alpha $-amides of $N$-terminal sites seem to be the aminoacid
residues, which are responsible for this transformation \cite
{ref2,ref8,ref53}. Specifically, in the alkaline Bohr effect the interaction
between the solvent and the $\beta 146His$ residue 
(a $C$-terminal histidine of $\beta $%
-chain) is one of the most probable reasons for the heme structure
modification and the rearrangement of neighboring aminoacid residues \cite
{ref2}. Our simulation confirms that the variation in $pH$ can result in
essential structural transformations in immediate proximity from the iron
atom. The strong $pH$-dependence for the times $\tau _{reb}$ and $\tau _s$
is a consequence of the transformations.

The ligand penetration from the solvent into the heme pocket is shown for $%
O_2$ or $NO$ in $Hb$ \cite{ref5} to be a process restraining the rate of
bimolecular recombination (\ref{eq9}). Therefore, the similarity between the 
$pH$-dependences for $\tau _{reb}$ and $\tau _s$ testifies that the change
in $pH$ has a slight effect on the oxygen migration in hemoglobin at the
mono- and bimolecular stages of recombination (\ref{eq9}). The weak $pH$%
-dependence for the obtained times of oxygen escape from the pocket confirms
this assumption. Such a $pH$-invariant behavior for the oxygen migration can
be interpreted by the independence of mobility for the hemoglobin side
chains (which seem to be responsible for the ligand transitions between
cavities of heme protein \cite{ref38}) on $pH$ of the solution.

\subsection{Efficiency of oxygen escaping from hemoglobin}

\begin{figure}
\psfig{file=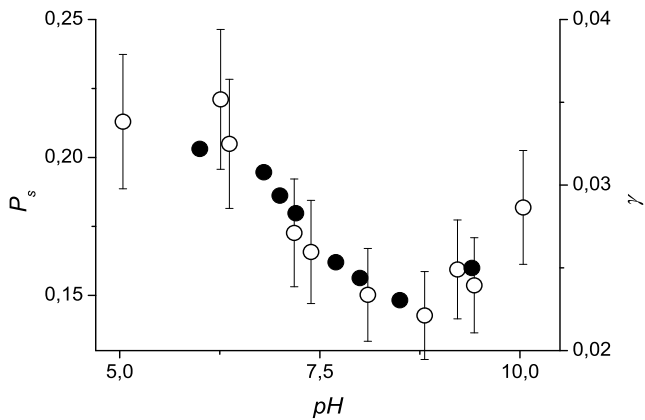,width=8.5cm,height=5.5cm}
\vspace{1 mm}
\caption{
The portion $P_s$ of the hemes remaining in unbound state after
realization of the geminate stage of $HbA$ reoxygenation (\ref{eq9}) 
(full circles -
our simulation) and the apparent quantum yield $\gamma $ of 
$HbA$ photodissociation ($\bigcirc$ - experiment 
\protect\cite{ref50,ref51}) depending on 
$pH$ values of the solution.}
\label{fig4}
\end{figure}

The obtained times $\tau _{reb}$ and $\tau _{esc}$ are used then in order to
estimate the efficiency of oxygen escape from hemoglobin as a function of 
$pH$. The efficiency is proportional to the quantum yield of photodissociation 
and can be associated with the portion $P_s$ of the hemes remaining in 
unbound state after realization of the geminate reoxygenation stage \cite{RefY}. 
In our
simulation the ratio $\tau _{reb}/\tau _{esc}$ falls within a range of
values from 0.4 up to 0.6. It implies that the portion $P_s$ can be determined 
from approximation(\ref{eq8}). 
The behavior of $pH$-dependence for the obtained quantity $P_s$
agrees well with one for the measured quantum yield of photodissociation 
\cite{ref51,ref51}. The quantity $P_s$ is shown in Fig. \ref{fig4} to be
proportional to the apparent quantum yield $\gamma $ for the whole
investigated scale of $pH$.

Notice that for the studied range of $pH$ values the quantity $C_s\tau
_{reb}/\tau _{esc}$ is considerably low in comparison with 1 ($C_s\tau
_{reb}/\tau _{esc}\sim 0.15\div 0.25$) and the time $\tau _{esc}$ is
practically constant. Therefore the relation of the portion $P_s$ with the
time $\tau _{reb}$ is close to the linear law:
\begin{equation}
\label{eqPs}P_s=\frac{C_s\tau _{reb}}{\left( \tau _{esc}+C_s\tau
_{reb}\right) }\approx C_s\frac{\tau _{reb}}{\tau _{esc}}\propto \tau _{reb}.
\end{equation}
In our study (see Fig. \ref{fig3} and Fig. \ref{fig4}) the $pH$%
-dependences obtained for $P_s$ and $\tau _{reb}$ are similar 
that testifies again
that the transport properties for the oxygen molecule in hemoglobin do not
depend on $pH$.

\subsection{Diffusion properties of oxygen migration in hemoglobin}

The analysis of X-ray diffraction data \cite{ref54,ref55} for oxygenated and
deoxygenated species of human hemoglobin (PDB ID $1HHO$ and $2HHB$,
correspondingly) shows that the cage radius $\rho $ to be associated with
the heme pocket size  is a quantity
ranging from 1 up to 5 \AA  (taking into account the Van der Waals
radiuses). Hence, the diffusion coefficient $D$ for the oxygen migration in
hemoglobin can be estimated from relation (\ref{eq7}): $D=\rho ^2/6\tau
_{esc}\sim 0.2\div 5\cdot 10^{11}$ m$^2$/s. This coefficient is intermediate
to diffusion coefficients for small molecules in water (~10$^{-9}$ m$^2$/s)
and solids (10$^{-18}$ m$^2$/s at $T<400K$) \cite{ref56}

According to the diffusion law, at a time instant $t_m$ the root-mean-square
displacement $\sqrt{\langle \left| \bf{r}(t)\right| ^2\rangle }$ of the ligand
from the iron is approximately equal to $\rho \sqrt{t_m/\tau _{esc}}$. It
implies that on the completion of kinetic measurements ($t_m=1.5$ ns \cite
{ref50}) the oxygen remains inside the protein and is localized in immediate
proximity from the heme pocket: 
$\sqrt{\langle \left| \bf{r}(t)\right| ^2\rangle }\approx 1.3\rho <7$ \AA . 
This conclusion is consistent with results of
spectroscopy investigation of motional dynamics 
for $CO$ in $Hb$ \cite{ref42}.

\section{Conclusion}

We have represented a simple model of the geminate recombination 
of heme
proteins with small ligands. The model takes into account dynamic properties
of ligand displacement in protein matrix. In the model the recombination is
due both to the heme-ligand rebinding and to the diffusion-like migration of
ligand between protein cavities. The temporal behavior for the probability $%
P(t)$ to find the heme in unbound state is specified in terms of two
parameters. They are the time $\tau _{reb}$ of heme-ligand rebinding for the
ligand inside the heme pocket and the time $\tau _{esc}$ of ligand escape
from the pocket.

We have applied our model in order to analyze 
a postphotodissociative geminate
reoxygenation of human hemoglobin at various $pH$ values of the solution. The
measured kinetic curves and the efficiency of oxygen escape from the 
hemoglobin
are well reproduced in our simulation. It testifies that the non-exponential
behavior for the probability $P(t)$ can be explained by a diffusion-like
migration of ligand over protein cavities. This conclusion is consistent
with recent kinetic measurements \cite{ref57}. We believe that the
theory-experiment agreement may be considered as an additional validation
for the glass-like model of proteins.

Our study demonstrates also that the variation in $pH$ can result in
considerable changes for the rate of heme-ligand rebinding. At the time, the
oxygen migration in hemoglobin depends slightly on $pH$. We have interpreted
this effect as a result of essential structural transformations in immediate
proximity from the iron atom. Certainly, this conclusion demands a more
detailed and thorough examination. In any case we suppose that the $pH$%
-induced modification of the initial stage of GR (if the modifications are
observed) can be explained by a change in the rate of heme-ligand rebinding.

\end{document}